\documentclass[proceedings]{rmaa}

\newcommand{\etal}{et~al.}

\newcommand{\HST}{{\em HST\/}}

\def\msun{\ifmmode {\rm M_\odot} \else M$_\odot$\fi}

\def\msunyr{\ifmmode {\rm M_\odot~yr^{-1}}\else${\rm M_\odot~yr^{-1}}$\fi}
\def\lam{\ifmmode {\lambda} \else {$\lambda$} \fi}
\def\lalpha{\mbox{Ly$\alpha$}}
\def\muobs{\ifmmode {\mu_{o}} \else  $\mu_{o}$ \fi}
\def\teff{\ifmmode {T_{eff}} \else $T_{eff}$ \fi}
\def\ilam{\ifmmode {I_\lambda} \else  $I_\lambda$ \fi}
\def\inu{\ifmmode {I_\nu} \else  $I_\nu$ \fi}
\def\fnu{\ifmmode {F_\nu} \else  $F_\nu$ \fi}
\def\tauh{\ifmmode {\tau_{\rm H}} \else $\tau_{\rm H}$ \fi}
\def\cm{\ifmmode {\rm cm} \else  cm \fi}
\def\cmmitwo{\ifmmode \rm cm^{-2} \else $\rm cm^{-2}$\fi}
\def\cmmithree{\ifmmode \rm cm^{-3} \else $\rm cm^{-3}$\fi}
\def\cmps{\ifmmode \rm cm~s^{-1}\else $\rm cm~s^{-1}$\fi}
\def\cmpsps{\ifmmode \rm cm~s^{-2}\else $\rm cm~s^{-2}$\fi}
\def\kmps{\ifmmode \rm km~s^{-1}\else $\rm km~s^{-1}$\fi}
\def\kmpspmpc{\ifmmode \rm km~s^{-1}~Mpc^{-1} \else
    $\rm km~s^{-1}~Mpc^{-1}$\fi}
\def\ergps{\ifmmode \rm erg~s^{-1} \else $\rm erg~s^{-1}$ \fi}
\def\ergpspcm{\ifmmode \rm erg~s^{-1}~cm^{-2}
  \else $\rm erg~s^{-1}~cm^{-2}$ \fi}
\def\ergpspcmphz{\ifmmode \rm erg~s^{-1}~cm^{-2}~Hz^{-1} \else $\rm
erg~s^{-1}~cm^{-2}~Hz^{-1}$ \fi}
\def\ergpspcmpa{\ifmmode \rm erg~s^{-1}~cm^{-2}~\AA^{-1} \else $\rm
erg~s^{-1}~cm^{-2}~\AA^{-1}$ \fi}
\def\ergpsphz{\ifmmode \rm erg s^{-1} Hz^{-1} \else
   $\rm erg s^{-1} Hz^{-1}$ \fi}
\def\mdoto{\ifmmode \dot M_0 \else $\dot M_0$ \fi}
\def\eg{e.g.}

\def\cf{cf.}

\title{The Puzzle of The Lyman Continuum Polarization of QSOs}

\author{Gregory A. Shields\altaffilmark{1},
        Eric Agol\altaffilmark{2}, and
        Omer Blaes\altaffilmark{3}
        }

\altaffiltext{1}{Department of Astronomy, University of Texas at Austin}
\altaffiltext{2}{Department of Astronomy, Johns Hopkins University}
\altaffiltext{3}{Department of Physics, University of California, Santa
Barbara}

\fulladdresses{
\item Gregory A. Shields: Dept. of Astronomy,
      Univ. of Texas, Austin TX 78712
     (shields@astro.as.utexas.edu)
\item Eric Agol: Dept. of Physics and Astronomy,
      Johns Hopkins University,
       Baltimore, MD 21218
      (agol@pha.jhu.edu)
\item Omer Blaes: Dept. of Physics,
      University of California,
      Santa Barbara, CA 93106
      (blaes@physics.ucsb.edu)
  }

\shortauthor{G. A. Shields et al.}
\shorttitle{Lyman Continuum Polarization of QSOs}

\listofauthors{G.~A.~Shields, E.~Agol, \& O.~Blaes}
\indexauthor{G.~A.~Shields}
\indexauthor{E.~Agol}
\indexauthor{O.~Blaes}

\ReceivedDate{2000 June 6}
\AcceptedDate{2000 June 6}

\resumen{  }
\abstract{
Recent spectropolarimetry of QSOs has revealed a surprising rise
in polarization in the Lyman continuum of several objects.
We discuss several recent attempts to interpret this feature,
including the role of Lyman limit absorption in PG 1222+228.
We present new theoretical results involving electron scattering
in a hot corona or wind above an accretion disk, and polarization
resulting from the relativistic returning radiation.
These mechanisms can lead to potentially observable
polarization at short wavelengths, but neither has quantitative
success in fitting the observed Lyman continuum polarization rises.
A renewed capability for ultraviolet spectropolarimetry from
space is urgently needed to provide further clues to the nature
of this phenomenon.
  }


\begin{document}

\maketitle

\section{Introduction}
\label{sec:Introduction}

Accretion onto black holes is widely believed to power active galactic
nuclei (AGN).  The accretion flow likely takes the form of an orbiting
disk, which gives efficient energy production and defines a natural axis
for jets and double radio sources. Specific observational confirmation of
the presence of disks in AGN has, however, been elusive.

Disks in AGN are defined by the black hole's mass, $M$, its dimensionless
angular momentum, $a_*$, and the accretion rate, $\dot M$.  For much of the relevant
range of parameters, the disk should be geometrically thin  and optically thick. 
In this case, the disk may emit much of its power as thermal emission in the
optical and ultraviolet.  Observations of QSOs 
indeed show a strong, broad continuum
component, called the ``Big Blue Bump'', that is suggestive of disk emission
(e.g., Shields 1978; Malkan 1983). Much effort has gone into observations at
infrared to X-ray wavelengths, polarization studies, and related theory, in
an attempt to confirm that this emission does indeed come from a thermally
emitting disk (see review by Koratkar \& Blaes 1999).  The case today remains
inconclusive.

One reason for the difficulty of proving the presence of disks in AGN is
the high orbital velocity of the material in the emitting zone.  This tends
to mask spectral features of a disk
that might have diagnostic value.  The bulk of the
thermal continuum should come from dimensionless radii $r_*
\equiv R/R_g \approx 10,$ where $R_g = GM/c^2$ is the gravitational
radius.  The corresponding orbital velocity is $\sim 0.3c$.  Detection of
highly broadened Fe K$\alpha$ emission has been cited as an argument for a
relativistic disk (Tanaka \etal\ 1995), but spectral features in the
optical and ultraviolet have not been identified.  One potential feature is
the Lyman edge of hydrogen, at rest wavelength 912~\AA.  Depending on the
parameters, this feature may be in emission or absorption or absent. It
seems likely to be present in some fashion at least in some objects, albeit
highly broadened by Doppler broadening and relativistic effects (cf.
Koratkar \& Blaes 1999).  In fact, only a small
fraction of QSOs, the so-called ``Lyman edge candidate QSOs'', show
indications of a broadened Lyman edge in absorption (\eg, 
Antonucci, Kinney, \& Ford 1989; Koratkar, Kinney,
\& Bohlin 1992).

The hot, low density gases in the atmosphere of an AGN disk should be
highly ionized.  Electron scattering should should then
give significant plane polarization.  A drop in polarization in the Lyman
continuum was predicted by Laor, Netzer, \& Piran (1990) because of the
increased absorption opacity from photoionization of hydrogen.  With this
motivation, spectropolarimetry was obtained with the  {\em Faint Object
Spectrograph} on the {\em Hubble Space Telescope} 
(\HST) for a number of QSOs (Impey \etal\ 1995; Koratkar \etal\
1995).  For several candidate Lyman edge QSOs, a surprising rise in
polarization was found.  From values of only $\sim 1\%$, the
polarization rises rather steeply around rest wavelength
$\sim 750$ \AA, reaching $\sim 5\%$ in PG1222+228 and $\sim 20\%$ in PG
1630+377.  Weaker polarization rises at a similar wavelength were found in
several other QSOs.  Until recently, these polarization rises appeared to be
associated only with candidate Lyman edge QSOs (Koratkar \etal\ 1998).

Lyman continuum polarization rises in QSOs have generated considerable
theoretical interest.  An interesting stellar atmosphere effect was
proposed by Blaes \& Agol (1996), in which the combined effects of electron
scattering and photoionization opacity give a polarization rise somewhat to
the short wavelength side of the Lyman edge.  This occurs for just the
range of disk effective temperatures, \teff, that should give a
Lyman edge in absorption in the total flux.  This process did not give
polarizations as high as 20\%, however.  Moreover, Shields, Wobus, and
Husfeld (1998, SWH) found that relativistic effects caused an additional
blueshift of the polarization rise, relative to the wavelength in the rest
frame of the orbiting gas.  The observed polarization rise then would occur at
a wavelength too short to fit the observations.  SWH noted that if the
polarization rises abruptly at the Lyman edge in the rest frame of the gas,
relativistic effects give a good fit to the observed wavelength dependence
of the polarization rise in PG1630+377.  The fit provides a possible
measure of the black hole spin.  However, no physical reason for the
postulated polarization rise is known.   Recently, Beloborodov
\& Poutanen (1999, BP) have offered an explanation for the polarization
rises in terms of electron scattering in a corona or wind above the disk.
In a different approach,  Lee \& Blandford (1997) discussed
scattering by resonance lines of heavy elements as a possible cause of the
Lyman continuum polarization rises.  However, they did not give a 
quantitative model of the observed polarization rises.

We report here theoretical investigations of several aspects of Lyman
continuum polarization rises in QSOs.  In \S 2, we summarize recent results
on PG1222+228, concerning the apparent coincidence of its polarization rise
with an intervening Lyman limit system (LLS).  In \S 3, we
present new results for electron scattering winds and coronae.  In \S 4, we
discuss the relevance of relativistic returning radiation to the Lyman
continuum polarization rises.  Our conclusions are summarized in \S 5.

\section{The Strange Case of PG 1222+228}
\label{sec:1222}

PG 1222+228 is a radio quiet QSO of apparent magnitude $B\approx 15.5$
(Schmidt \& Green 1983).  Impey \etal\ (1995) observed a polarization rise
at $\lambda \approx 750$ \AA\ that coincides with a sharp drop in flux
(Figure 1).  They suggested that this was a coincidental LLS,
corresponding to an observed narrow absorption line system at $z$ = 1.486.
Shields (2000) considered the possibility that the agreement in wavelength
between the polarization rise and the absorption feature was not a
coincidence. Broad absorption line (BAL) QSOs often show a rise in
polarization in the BAL troughs (Schmidt \& Hines 1999; Ogle \etal\ 1999;
and references therein), reaching values as high as $\sim 8$ to 10 percent
from $\sim 1$ outside the troughs.   This is explained by a geometry in
which the polarized flux results from a scattering source (perhaps an
electron scattering wind) that is larger than the continuum source.  The
outflowing gas producing the absorption troughs blocks the line of sight to
the continuum source but not the scattering source.  In the troughs, one
sees mainly the less luminous, but highly polarized scattering source.  Could
a similar geometry explain the polarization rise in PG 1222+228?

\begin{figure}
  \begin{center}
    \leavevmode
    \includegraphics[width=3in]{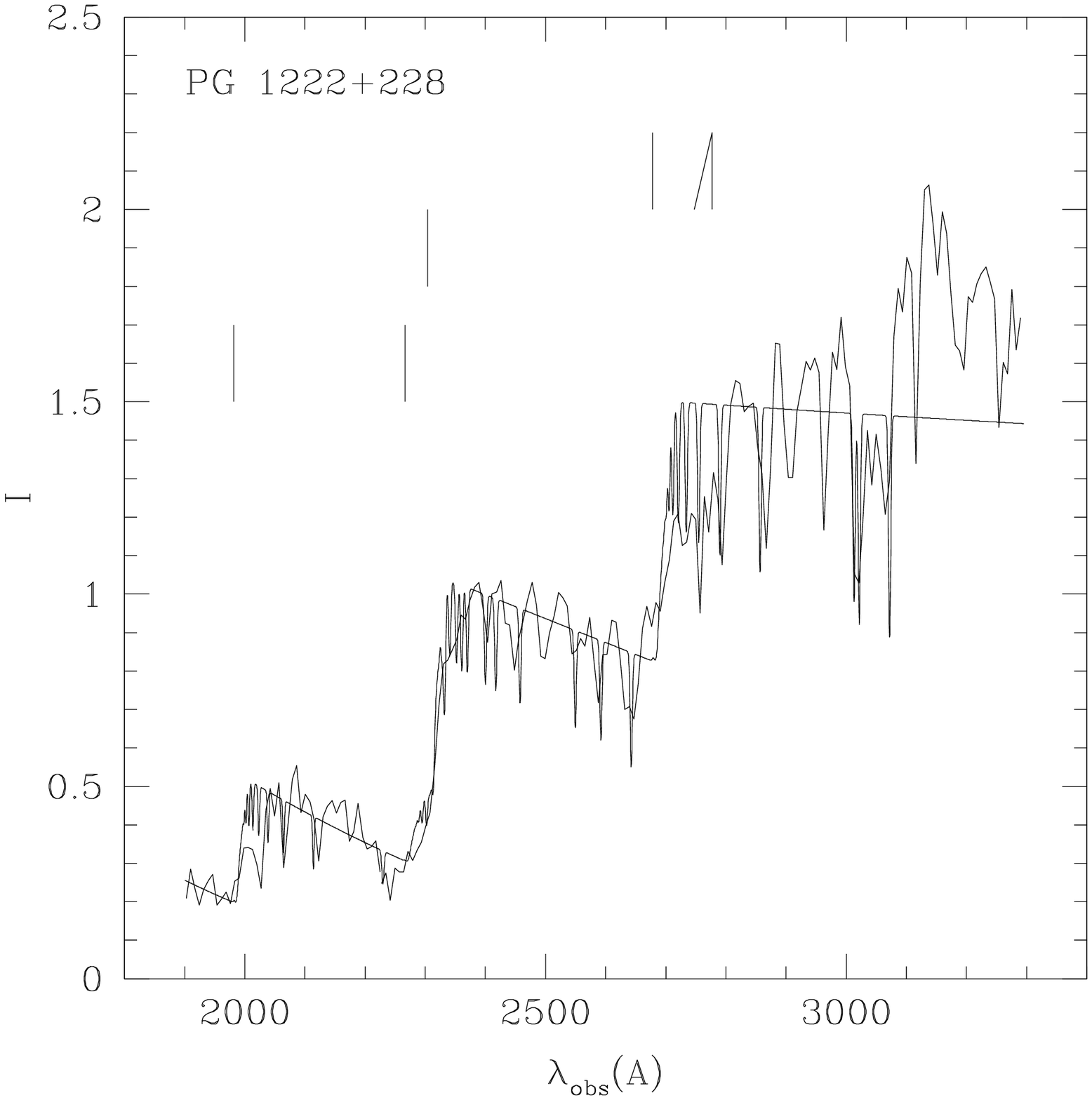}
	\hspace{5mm}
    \includegraphics[width=3 in]{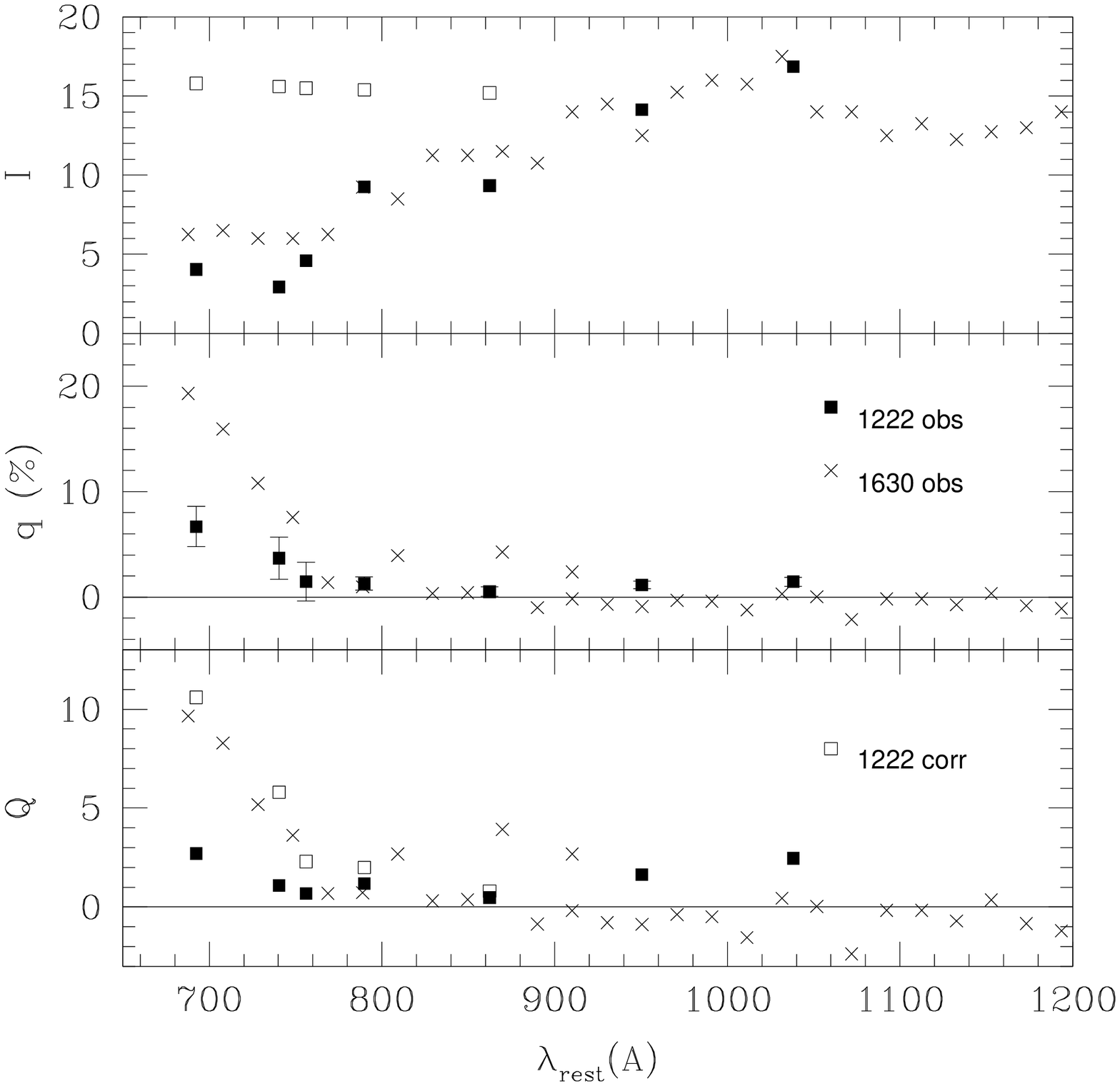}
    \caption{
      \HST spectrum of PG 1222+228 (Impey \etal\ 1995) fit with a
      model involving several Lyman limit systems.  See text and
      Shields (2000) for details.
      Reproduced from Shields (2000).  Copyright 2000,
      Astronomical Society of the Pacific; reproduced with permission.
      }
    \caption{
      Polarized flux of PG 1222 corrected for Lyman limit absorption.
      Note rise in polarized flux in the Lyman continuum, resembling
      the case of PG 1630+377 (Koratkar \etal\ 1995).
      Adapted from Shields (2000).  Copyright 2000,
      Astronomical Society of the Pacific; reproduced with permission.
      }
  \end{center}
\end{figure}

The flux drop in PG 1222+228 is close to the wavelength of Ne
VIII $\lambda$775, observed in some BAL QSOs
(\eg, Arav \etal\ 1999).   However, the
feature in PG 1222+228 does not recover with decreasing wavelength within a
normal velocity range.  Expected BALs such as C IV and N V
are not present, and the soft X-ray emission of PG 1222 +228 is stronger
than is typical for BAL QSOs (Shields 2000).  
Alternatively, Shields (2000) considered the
possibility of an intrinsic, high velocity LLS.  Such a feature would be
unprecedented.  However, Richards \etal\ (1999) argue that a large fraction
of the narrow, high velocity absorption line systems in luminous QSOs,
usually assumed to be intervening, are in fact intrinsic.  In
order to obscure the continuum source but not the scattering source, the
outflowing gas would be quite close to the central source.  In this case,
the lack of measurable change in velocities in the associated narrow
lines, over a period of seven years, argues against this picture.

Reverting to the intervening LLS interpretation, Shields (2000) showed
that a good fit to the flux drop in PG 1222+228 was given by LLS associated
with observed absorption line systems at z= 1.486 and 1.524 (see
Figure 1).  The fit is supported by the recovery of the continuum below the
edge in a manner consistent with the $\nu^{-3}$ dependence of the optical
depth and a typical intrinsic power-law continuum $L_{\nu} \sim
\nu^{-1.8}.$  Two additional LLS occur at z = 1.174 and 1.938.  The quality
of the fit and the straightforward nature of this explanation support the
conclusion that the Lyman continuum polarization rise in this QSO is a
coincidental feature that happens to agree in wavelength with the LLS at $z$ =
1.486, 1.524 .

This result implies that the observed, polarized flux should be corrected
for the intervening absorption.  Figure 2 shows the corrected Stokes flux,
$Q^\prime_\lambda$, rotated to the mean position angle of 168 degrees.
($Q^\prime$ has the advantage of avoiding the positive bias of the
polarization in data with low signal-to-noise, as discussed by Koratkar \etal\
1995.)  This quantity rises strongly with decreasing wavelength in the
region of the polarization rise.  This behavior resembles the case of PG
1630+377, and it poses especially severe demands on theoretical models.  The
harder ultraviolet flux after correction for absorption requires a higher
effective temperature for the accretion disk atmosphere.  Shields (2000)
found a reasonable fit for $a_* = 0.5, M_9 = 8.8,$ and \mdoto = 86.  These
parameters give a luminosity $L = 0.36L_{Edd},$ barely compatible with a
geometrically thin disk.

A consequence of this analysis is that PG 1222+228 may not be a
true Lyman edge quasar.  All known polarization rise QSOs have heretofore
been associated with candidate Lyman edge QSOs (Koratkar \etal\ 1998).
Correspondingly, the presence of a Lyman edge in absorption in the disk
continuum has been a feature of several proposed models for the Lyman
continuum polarization rises.  The existence of a polarization rise QSO
without a Lyman edge in the total flux would be a challenge for such models.

\section{Winds and Coronae}
\label{sec:winds}

BP have investigated an explanation for the Lyman continuum polarization
rises that involves a hot corona or wind overlying the accretion disk.
Coronae above AGN disks have been considered before in various contexts,
although the heating mechanism is uncertain (\eg, Haardt \& Maraschi 1991).
In the case of a static corona, BP assume
a plane parallel geometry with an underlying photosphere that emits a black body
continuum at $kT = 3$~eV and a sharp Lyman edge in absorption.
For a Thomson optical depth
$\tau_{\rm T} = 1$, a substantial number of photons undergo more than one
scattering before escaping from the top of the corona.   The
polarization of a scattered photon depends on its direction going into the
last scattering.  For photons scattered twice or more, the source of the
photons for the last scattering is the corona itself. Therefore, the
source is effectively limb brightened.  This results in a substantial
degree of
polarization, parallel to the disk axis (positive in our convention).  For
frequencies below the Lyman edge, the flux is dominated by the primary
emission together with singly scattered photons, and the polarization is
essentially the perpendicular (negative) polarization of an electron
scattering atmosphere (Chandrasekhar 1960).  Just above the Lyman edge, the
scattered flux is dominated by singly scattered photons. At frequencies
substantially above the Lyman limit, however,
most of the observed flux consists of
multiply scattered photons, and this leads to an increasing, parallel
polarization.  This model naturally gives an increase in polarization in the
Lyman continuum, rising toward shorter wavelengths.

The observed flux drop and polarization rise become more gradual as the
coronal temperature increases (BP's Figure 1).  Indeed, Hsu and Blaes
(1998) calculated the polarization in a similar geometry with $kT \sim 100$
keV, and found a gradual increase in polarization to values $\sim 10 -
15\% $ at photon energies above 1 keV.  BP find a polarization rise to
nearly 20\% at 600~\AA,  resembling the observed polarization rise in PG
1630+377.

We have made independent calculations of the polarization in the BP model,
using a code based on the iterative scattering method of Poutanen \& Svensson
(1996).  Our
results confirm the behavior of the flux and polarization found by BP for
their assumed parameters.  However, a plot of the polarized flux is
revealing (Figure 3).   Following Koratkar \etal\ (1995), we give the
observed, polarized flux in terms of the rotated Stokes flux, $Q^\prime.$
The observed polarized flux is essentially
zero redward of the Lyman limit, and rises steeply at about
750~\AA.  The model's polarized flux shows a more gradual rise.  Redward of
the Lyman limit, the model polarization is perpendicular to the
disk axis, as noted by BP.  Just below the Lyman edge, it crosses into
positive (parallel) values, and rises rather gradually before actually
dropping at wavelengths below $\sim$750~\AA.  This is rather different
from the observed behavior.  Conceivably, some additional polarization
process, such as scattering by material at larger radius, just offsets the
negative polarization at longer wavelengths. In Figure 3, we also show the
resulting polarization if a wavelength independent polarization of 2.5\% is
added to the model, resulting in zero polarization at longer wavelengths.
The polarized flux is computed from the resulting polarization and the total
flux.  The polarized flux rise is still too gradual and fails to reach the
observed values at the shortest wavelengths.

\begin{figure}
  \begin{center}
    \leavevmode
    \includegraphics[width=3in]{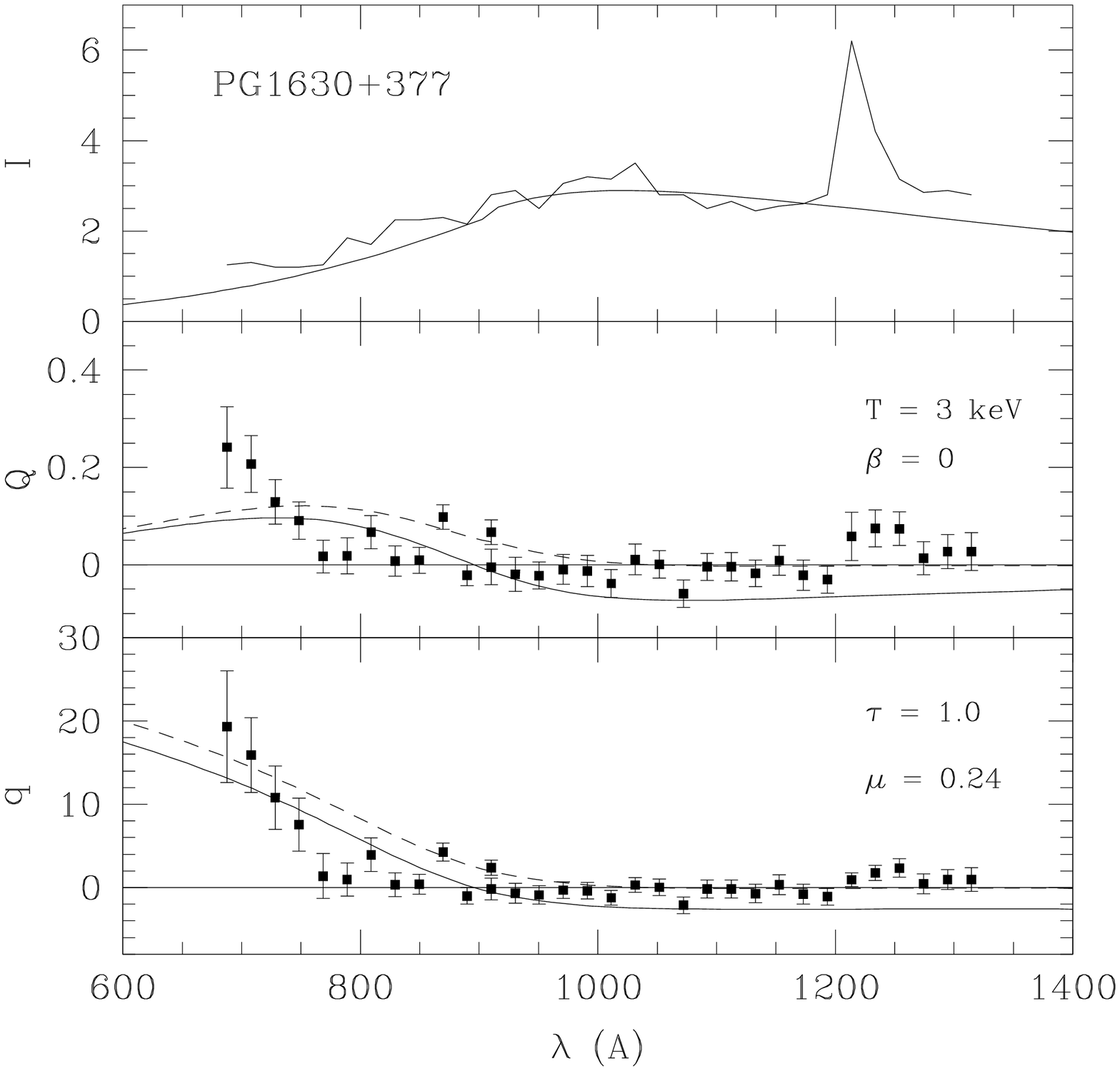}
        \hspace{5mm}
    \includegraphics[width=3in]{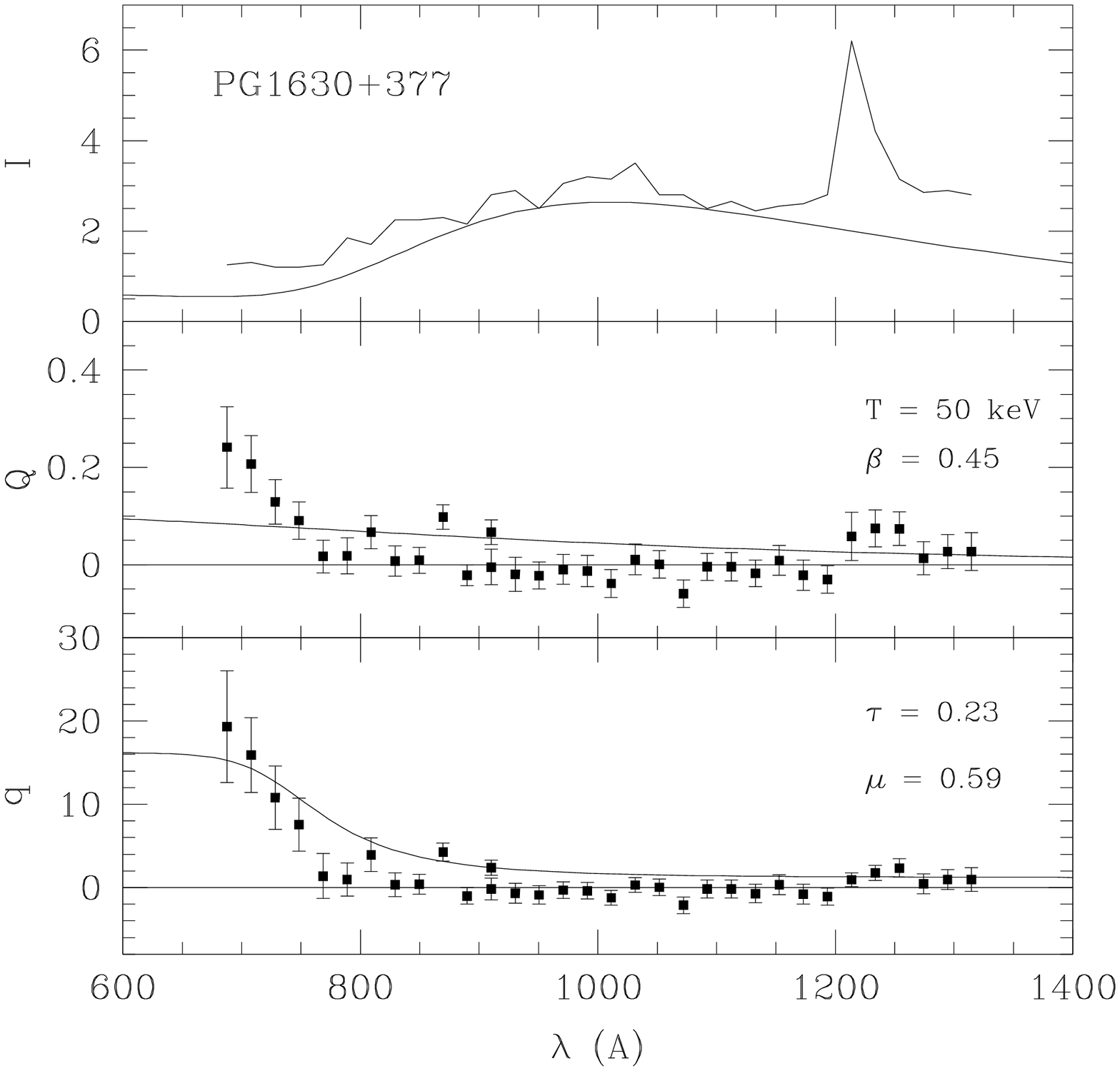}
    \caption{
      Polarization versus wavelength in static corona overlying
      accretion disk.  Solid line shows model results; dashed
      curve shows model results plus an added wavelength 
      independent polarization of 2.5\%.  See text for details.
      Observations from Koratkar \etal\ (1995).
      }
    \caption{
      Polarization versus wavelength in electron scattering wind overlying
      accretion disk.  See text for details.
      }
  \end{center}
\end{figure}

BP reject the static corona model on the basis of the change in sign of
polarization around the Lyman limit, which is not observed.  They also note
that in this model, a strong Lyman edge in absorption would be seen for
relatively face on viewing angles, and such edges are rarely if ever in
observed in QSOs.  Our results for the polarized flux further strengthen the
case against a static corona as a cause of the Lyman continuum polarization
rise in PG 1630+377 and similar objects.

BP find more promise in a model involving a corona with a mildly
relativistic outflow velocity.  Beloborodov (1998) showed that, for an
outflow velocity $\beta \equiv v/c = 0.45$ (as might occur for an
e$^\pm$ plasma accelerated by the radiation pressure of the disk emission),
the scattered radiation can be strongly polarized parallel to the disk
axis.  This results from the fact that the disk emission is effectively limb
brightened by relativistic aberration in the rest frame of the scattering
material.  BP performed single-scattering, frequency dependent calculations
with allowance for a high wind temperature and resulting shift of the
frequencies of scattered photons.  For the primary disk continuum, BP
assumed a toy continuum that simulates a Lyman edge in absorption smeared
out by relativistic effects near the black hole.  For a wind optical depth
of
$\tau_{\rm T}$ = 0.23 and $kT = 50$ to 100 keV, the polarization rises in a way
resembling that observed in PG 1630+377.  The polarization rise largely
results from the dropping flux below the Lyman edge, whereas the polarized
flux, resulting from the hot, moving corona, does not show the Lyman edge.

We have modified our iterative scattering code to reproduce the wind model of
BP.  Consider a plane parallel, vertical wind with constant Lorentz factor
$\gamma=(1-\beta^2)^{-1/2}$.  The radiation field is time independent
in the lab frame, but spatial gradients in this frame will mean that it
will not be time independent in the comoving frame.  
Using standard Lorentz invariants (e.g.  Mihalas \& Mihalas 1984)
we may write the lab frame radiative transfer equation
in terms of comoving frame quantities, viz.
\begin{equation}
\mu{\partial {\bf I}_\nu\over\partial\tau_{\rm T}}=(1-\beta\mu)
\left[{\sigma_{\rm CS}(\nu_{\rm c})\over\sigma_{\rm T}}{\bf I}_\nu
-{1\over\gamma^3(1-\beta\mu)^3}{\bf S}_{\rm c}\right].
\end{equation}
Here ${\bf I}_\nu$ is the lab frame specific intensity Stokes vector;
$\sigma_{\rm CS}(\nu_{\rm c})$ is the thermal Compton scattering cross
section, evaluated at the comoving frame frequency
$\nu_{\rm c}=\gamma(1-\beta\mu)\nu$;
and ${\bf S}_{\rm c}$ is the comoving frame polarized source function,
evaluated from the comoving frame intensity at the comoving frame frequency
and direction cosine $\mu_{\rm c}=(\mu-\beta)/(1-\beta\mu)$.
Expressions for $\sigma_{\rm CS}$ and ${\bf S}_{\rm c}$ may be found in
Poutanen \& Svensson (1996).

We iteratively solved the lab frame transfer
equation in the same way as in the static corona model.  For
$\beta = 0.45, kT = 50~$keV, and $\tau_{\rm T} = 0.23$,
we obtained good agreement with BP's predicted flux and
polarization as a function of wavelength (Figure 4).  (BP used a single
scattering approximation, which is a good approximation at these wavelengths
and low optical depths.  We follow BP in using a plane parallel
wind truncated at some optical depth, $\tau_{\rm T}$, as a crude 
approximation to a wind that presumably diverges at some height comparable
with the radius, above which there is little further optical depth.)
Once again, however, the polarized flux seems to
provide a more vivid test of agreement with observation.  Figure 4 shows that
the polarized flux rises much more gradually, with decreasing wavelength, than
is observed.  The model thus appears to have difficulty giving a quantitative
fit to the data for PG 1630+377.  In retrospect, this conclusion may 
seem unsurprising.  The polarized flux is a direct measure of the scattered
radiation, and since the electron scattering cross section does not
change across the Lyman edge, neither should the polarized flux change.

Both the static corona and wind models rely on the presence of a deep Lyman
edge in absorption to give a rise in polarization in the Lyman continuum.
The results described above for PG 1222+228 suggest that this object may not
be a Lyman edge quasar, once correction is made for the intervening LLS
absorption.  In this case, coronal scattering models may face a further
difficulty.

\section{Returning Radiation}
\label{sec:returning}

Radiation leaving the surface of a disk near a black hole will suffer
various relativistic effects as it propagates away from the point of
emission.  These include the relativistic Doppler effect, the gravitational
redshift, the relativistic aberration, and general relativistic bending of
the light path.  The bending causes some light, mainly from the innermost
parts of the disk, to return to the equatorial plane and strike the disk,
often at a larger radius.  This ``returning radiation'' is minor for disks
around nonrotating black holes but can be substantial for rapidly rotating
holes (Cunningham 1976).  This is true because a larger black hole
angular momentum gives a smaller value of the radius $r_{ms}$ of the innermost
stable circular orbit, usually taken to be the inner edge of the disk.
For prograde disks around rapidly rotating black holes with
$a_* = 0.9978$, the
returning radiation flux can amount to some 20\% of the locally generated
flux at larger radii (Cunningham 1976).  Because polarization of light
emitted from a scattering atmosphere depends strongly on the directionality
of the radiation going into the last scattering, the returning radiation
might have a substantial effect on the polarization of the total radiation
from the disk.

This process has been considered recently by Agol \& Krolik (2000) in the
context of the inner boundary conditions for
disks around black holes.  Krolik (1999) and Gammie (1999) have discussed
the possibility that plunging material inside $r_{ms}$ exerts a magnetic
torque on the material just outside $r_{ms}$, causing an increase in the
angular momentum flow outwards through the disk and the associated local
dissipation of energy.  Agol \& Krolik discuss several implications of this
process, including the equilibrium black hole spin, the disk energy
distribution, and polarization.  They parametrize the effect of the inner
torque in terms of $\Delta\epsilon$, the increase in efficiency of energy
production per unit mass accreted, where $\epsilon \equiv L/\dot{M} c^2$,
compared with zero inner torque.
They trace the trajectory of the emitted photons using a numerical method
described by Agol (1997).  The effect of a strong inner torque is to
increase dramatically the flux emitted by the innermost radii in the disk,
just outside $r_{ms}$.  This radiation is strongly affected by
gravitational bending; and it can increase greatly the amount of returning
flux striking the disk at larger radii.  For $a_*$ near unity, a large
fraction (up to 58\%) of the extra flux resulting from the inner torque
returns to the disk. Agol \& Krolik (2000) raised the possibility that the
polarization resulting from returning radiation might be relevant to the
Lyman continuum polarization rises.

We consider here this possibility in the context of the toy model described
by SWH.  In their model, a given point on the disk emits
a black body continuum at the local effective temperature for  $\nu < \nu_H;$
 for higher frequencies, it emits as a black body at $T =
0.8409T_{eff}$.  This crudely simulates a Lyman edge in an LTE atmosphere
with a large Lyman continuum opacity.  However, whereas SWH postulated a sharp
increase in polarization at the Lyman edge, we assume here that the emitted
radiation has the polarization of an electron scattering atmosphere at all
frequencies.  Our purpose is to explore whether relativistic returning
radiation can give a polarization rise at the Lyman edge in the radiation
observed from the disk as a whole.  This might occur because the drop in
brightness temperature in the Lyman continuum will cause the Lyman continuum
to be emitted from smaller radii in the disk than the radiation just redward
of the Lyman limit.  The radiation observed will have a larger contribution of
scattered, returning radiation in the Lyman continuum than at longer
wavelengths.  The scattered returning radiation, becoming more important
just at the Lyman limit, could give an abrupt rise in polarization.

SWH gave a model for PG
1630+377 with $a_* = 0.5$, $M_9 \equiv M/10^9~\msun= 5$,  and $\mdoto \equiv
\dot{M}/1~\msunyr = 27$, giving a maximum disk effective temperature of
$38,000$~K.  Using the numerical method of Agol (1997), we computed the
polarization, including scattering of the returning radiation according to
the diffuse reflection law of Chandrasekhar (1960).  Figure 5 shows that the
polarization of the
 observed radiation is affected insignificantly by the returning radiation,
except at very high frequencies that carry little flux.  The weak effect of
returning radiation is not surprising, since for a moderate value of $a_*$
and no inner torque, there is little returning radiation.  We have
calculated a model of the same nature, still with $\Delta\epsilon = 0$, but
with $a_* = 0.9978$.  This model also has $\mdoto = 27$, but it has $M_9 =
60$, so as to give again a maximum effective temperature of 38,000 K.
(This is a simple device to preserve a substantial Lyman edge in the
local emission and a rough fit to the observed energy distribution.)  Figure
5 shows that there is now a significant degree of polarization in the Lyman
continuum region, reaching about 6\% at observed frequency of log $\nu =
16.1$.  However, this falls well short of the observed 20\% polarization
of PG 1630+377; and the polarization rise is too gradual to explain the
observed rises in PG1630+377 or PG 1222+228.  (In order to show a
wider range of frequencies, we use a log $\nu$ scale in Figure 5.  However,
inspection of Figures 3 and 4 suffices to show that the observed
polarization rise is much more abrupt than predicted by this model.)
As noted by SWH, for a
rapidly rotating black hole and the required disk effective temperature,
the Lyman continuum is emitted at small radii.  This emission suffers strong
relativistic effects that give a gradual rise in the
observed polarization, reaching a maximum at wavelengths much to the blue of
the Lyman limit.    As a further example, we considered the case
$a_* = 0.9978$ and
$\Delta\epsilon = 1$, so that most of the disk's luminosity results from the
inner torque and comes from small radii.  This should give the maximum
effect of returning radiation.  In order to preserve $T_{max} = 38,000$~K,
this model had $M_9 = 60$ and
$\mdoto = 1.6$.  Figure 5 shows that the model has strong polarization at
high frequencies, reaching roughly 25\% at a frequency of log $\nu$ =
16.3.  Again, the polarization rise is much too gradual to the fit the
observations.  There is a large rotation of the position angle of the
polarization resulting from general relativistic effects (see also Figure 11
of Agol and Krolik 2000).

\begin{figure}
  \begin{center}
    \leavevmode
    \includegraphics[width=3in]{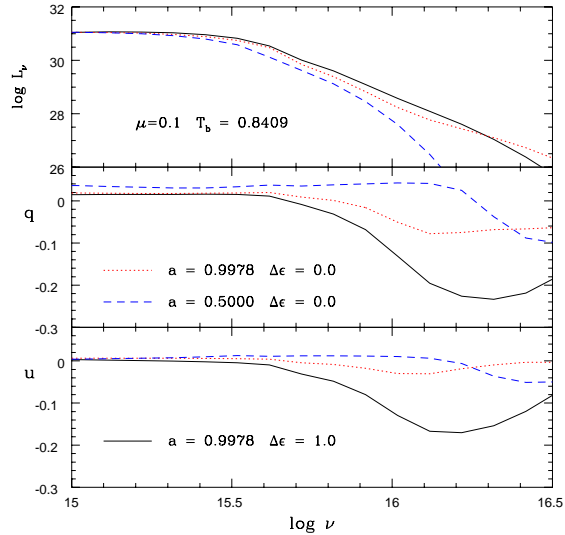}
    \caption{
      Polarization of thermal emission from accretion disk 
      including returning radiation.  See text for details.
      }
    \label{fig:returning}
  \end{center}
\end{figure}

As noted by SWH, a moderate black hole spin is required to
avoid excessive blueshifting and smearing of the polarization rise.
However, returning radiation then is insufficient to give polarization
rises of the observed magnitude.  This dilemma seems likely to doom any
attempt to use relativistic returning radiation to explain the presently
known instances of a Lyman continuum polarization rise.  The
strong polarization resulting from returning radiation might nevertheless be
observable in some QSOs.  In general, it will occur for disks around rapidly
rotating black holes, generally at high frequencies where the flux is rapidly
dropping.  Only at frequencies above the black body peak for the
hottest radii in the disk will the emission be dominated by the innermost
radii that give strong returning radiation.

\section{Conclusions}
\label{sec:conclusions}

We have explored several models to explain the Lyman continuum
polarization in QSOs.  Neither Compton scattering nor
returning radiation appear to be capable of explaining the
strong, rapid rise in polarized flux observed for PG 1630+377.
The same seems likely to be true for PG 1222+228, especially
after correction for the LLS absorption.

Quite apart from the issues discussed above, there is the 
observed polarization of the \lalpha\ line in PG 1630+277.
It is polarized in a similar positional angle to the 800 \AA\
continuum polarization, to a degree of about 3\%
(Koratkar \etal\ 1995).  The wavelength
of the feature in polarized flux actually is on the red wing of
the emission line in total flux, roughly at the expected
wavelength of the N V $\lambda$1240 resonance line (\cf\ Shields 2000).
Certainly, the emission line in the polarized flux does not
share the large blueshift of the polarization rise in the
continuum, if it is indeed associated with the Lyman edge of
hydrogen.  Discussions of the Lyman continuum polarization rise
have tended to ignore the ``\lalpha'' feature, although Koratkar
\etal\ (1995) did suggest scattering of optically thin hydrogen
emission followed by passage through an absorbing layer of
modest optical depth at the Lyman limit.  From another point
of view, however, there are only two natural position angles
in the context of a disk geometry (parallel and perpendicular
to the disk axis).  The polarized Lyman continuum and Lyman alpha
could come from two different sources, which might coincidentally
have the same position angle.  The polarized \lalpha\ feature
has a normal broad emission-line width, 
and presumably comes from material
at a radius $\sim10^4 R_g$.  The polarized Lyman continuum may
come from a smaller radius, as in the models of SWH and Blaes
and Agol (1996).  Here, for a thermal intrinsic line width,
thermalization may suppress the \lalpha\ emission, suggested
by simple estimates based on the black body limit.

From a broader perspective,  most proposals for the Lyman continuum
polarization rise seem contrived to explain this particular
observation, rather than following naturally from a more comprehensive
theory of AGN.  Thus, they do not address the issue of the
statistical incidence of the polarization rise phenomenon.
Unfortunately, there is currently no observational capability
to confirm and extend the measurements of Lyman continuum
polarization in QSOs.  The \lalpha\ forest is an increasing
problem at higher redshifts, and there is an urgent need for
a renewed capability to do ultraviolet spectropolarimetry
from space.

\acknowledgments

The authors acknowledge useful communications with A. Beloborodov
and R. Antonucci.
G.A.S. gratefully acknowledges
support from the Texas Advanced Research Program 
under grant 003658-015 and the Space Telescope Science Institute
under grant GO-07359.02. 
E.A. and O.B. acknowledge support from NSF grants 
AST~96-16922 and AST~95-29230 respectively.
This work was carried out in part at the Institute for Theoretical Physics,
University of California, Santa Barbara, supported by NSF grant PH94-07194.


\end{document}